\documentclass[prb,preprint]{revtex4-1} 

\usepackage{amsmath} 
\usepackage{amsfonts} 
\usepackage{graphicx} 
\graphicspath{{graph/}}
\usepackage[usenames]{color}
\usepackage{amsmath,amssymb}
\usepackage{gensymb}
\usepackage{natbib}
\usepackage{xcolor}

\begin{document}


\title{Turbulent dispersion of breath by the wind}



\author{Florian Poydenot}
\author{Ismael Abdourahamane}
\author{Elsa Caplain}
\author{Samuel Der}
\author{Antoine Jallon}
\author{In\'es Khoutami}
\author{Amir Loucif}
\author{Emil Marinov}
\author{Bruno Andreotti}
\email{andreotti@phys.ens.fr}
\affiliation{Laboratoire de Physique de l'\'Ecole Normale Supérieure (LPENS), CNRS UMR 8023, \'Ecole Normale Supérieure, Université PSL, Sorbonne Université, and Université de Paris, 75005 Paris, France.}


\date{\today}

\begin{abstract}
The pioneering work of G.I.~Taylor on the turbulent dispersion of aerosols is exactly one century old and provides an original way of introducing both diffusive processes and turbulence at an undergraduate level. Light enough particles transported by a turbulent flow exhibit a Brownian-like motion over time scales larger than the velocity correlation time. Aerosols are therefore subjected to an effective turbulent diffusion at large length scales. However, the case of a source of pollutant much smaller than the integral scale is not completely understood. Here, we present experimental results obtained by undergraduate students in the context of the COVID-19 pandemic. The dispersion of a smoke of oil droplets by a turbulent flow is studied in a wind tunnel designed for pedagogical purposes. It shows a ballistic-like regime at short distance, followed by Taylor's diffusive-like regime, suggesting that the continuous cascade is bypassed. Accordingly, measurements show that the CO$_2$ concentration emitted when breathing typically decays as the inverse squared distance to the mouth, which is not the law expected for a diffusion. The experiment offers the possibility for students to understand the role of fluctuations in diffusive processes and in turbulence. The non-linear diffusive equations governing aerosol dispersion, based on a single correlation time, allow us to model the airborne transmission risk of pathogens, indoors and outdoors. We discuss the pedagogical interest of making students work on applied scientific problems, as the results obtained in this study have been used to provide public health policy recommendations to prevent transmission in French shopping malls.
\end{abstract}

\maketitle 

\section{Introduction} 
The impossibility of turning experimental physics courses online during the COVID-19 pandemic has motivated the following question: how can we use experimental physics projects to design rational methods to reduce the viral transmission risk indoors and keep this teaching going on during lockdowns? A group of students studying experimental physics at our university have therefore worked on the risk of airborne transmission of SARS-CoV-2, seen through complementary aspects: the dispersion of viral particles indoors and outdoors, reported here, the filtration efficiency of face masks, the calibration of CO$_2$ sensors using candles in a closed transparent box, the analysis of epidemic data, etc. This project has required a detailed understanding of both diffusion processes and turbulent flows.

However, turbulence is not an easy physics to teach. Fluid mechanics is mainly taught at undergraduate level through hydrostatics, potential laminar flows and viscous flows. The concept of turbulence appears when teaching the drag force exerted on a moving spherical solid, as a function of the Reynolds number (see section \ref{SectionReynolds}). In this approach, one of the fundamental aspect of turbulence, namely space and time fluctuations, is bypassed. Here we show that the dispersion of a passive scalar by a flow provides an accessible way of teaching turbulence, theoretically as well as through very simple experiments. In section \ref{sec:experiments}, we make use of controlled experiments performed both in a small wind tunnel used for pedagogical purposes, using a smoke of oil droplets, and in the large corridors of two shopping malls, under various ventilation conditions, using CO$_2$. We show that the concentration decays faster than expected by the turbulent dispersion theory of Taylor,\cite{taylor_diffusion_1922} following a ballistic law. Indeed, introducing a single velocity decorrelation time provides a perfect fit to the data, suggesting that the cascade is bypassed.\cite{villermaux_histogramme_1998,villermaux_geometry_1999,villermaux_short_2001,villermaux_mixing_2019} The Kolmogorov scaling law is not observed. The work presented here can be used to introduce students to the turbulent dispersion not only of pollutant molecules, but also of aerosols. We finally discuss the relevance of the results to the airborne transmission of SARS-CoV-2, which has motivated this research. SARS-CoV-2 is mainly transmitted through the airborne route:\cite{greenhalgh_ten_2021,morawska_it_2020} infection happens when a susceptible person inhales viral particles emitted by another infected person. In between their respective respiratory tracts, the air carrying viral particles is gradually diluted by turbulent dispersion. Near an emitter, the concentration of viral particles is therefore higher than far away, which has implications on the airborne transmission risk.

\section{Turbulent dispersion of particles}
\subsection{Motion of inertial particles dragged by a fluid}
\label{SectionReynolds}
The drag force on a spherical particle is a simple way of introducing different aspects of turbulence and in particular the concept of hydrodynamic regimes. Consider a solid particle of diameter $d$ moving at a constant velocity $\mathbf{v}$ with respect to a fluid at rest, of density $ \rho_f $ and viscosity $\eta$. These four parameters can be used to define characteristic scales of mass, length and time, leaving one dimensionless number controlling the flow around the particle: the particle-based Reynolds number $\mathcal{R}$, which compares inertial to viscous effects
\begin{equation}
\mathcal{R}=\frac{\rho_f v d}{\eta}=\frac{v d}{\nu},
\end{equation}
where $\nu=\eta/\rho_f$ is the kinematic viscosity.

In the viscous regime, at low Reynolds number $\mathcal{R}$, the drag force exerted by the fluid on the particle, which results from the distributed viscous stress, can be estimated from dimensional analysis. The velocity gradient scales as $v/d$. The force exerted by the fluid on the particle can then be written simply as the viscous stress $\sim \eta v/d$ times the surface $\sim d^2$, i.e.
\begin{equation}
\label{stokesdim}
\mathbf{f} \sim - \eta d\,\mathbf{v}.
\end{equation}
In these dimensional equations, the symbol $\sim$ is used to claim that the equation is valid up to a multiplicative factor which does not depend on any parameter.
For a rigid sphere, each surface element contributes equally to the resultant of hydrodynamic forces. The area of the sphere is equal to $\pi \, d^2$, the multiplicative factor in front of equation (\ref {stokesdim}) is $3 \pi$ so that the drag force, called Stokes force in this limit, reads\cite{ll_fluidmech}
\begin{equation}
\label{stokes}
\mathbf{f} = -3 \pi \eta d\,\mathbf{v}.
\end{equation}
In the frame of reference of the solid, the fluid is moving at a velocity $-\mathbf{v}$. In the viscous regime, the kinetic energy per unit volume $\frac 12\rho_f {v}^2$ is dissipated when a fluid particle approaches the solid surface.

At large Reynolds number $\mathcal{R}$, viscous tangential stress becomes negligible in front of pressure. By symmetry, the force $\mathbf{f}$ acting on a spherical particle is still colinear with the velocity $\mathbf{v}$ but no longer depends on viscosity. The main force comes from the pressure asymmetry between both sides of the particle. Indeed, when streamlines converge (upstream of the particle, in its frame of reference), turbulent fluctuations are damped while they are amplified when streamlines diverge (downstream of the particle). Kinetic energy far upstream of the particle is converted to pressure (internal energy) at the surface. The conservation of energy, called the Bernoulli equation, would lead to an overpressure equal to $\frac 1 2 \rho_f {v}^2$ at the surface of the particle. However, on the flanks of the particle, the shear boundary layer separates and forms a strongly dissipative recirculation bubble behind the particle. In the wake, turbulent fluctuations lead to an enhanced energy dissipation: kinetic energy is not converted into pressure. The pressure on the downstream face of the particle is roughly the pressure far from the particle. The total force is the product of the pressure by the surface
\begin{equation}
\mathbf{f} = -\frac{\pi}{8} C_\infty \rho_f d^2\, v \mathbf{v}.
\label{eq:dragturb}
\end{equation}
The multiplicative factor $C_\infty$ is called the drag coefficient, and depends on the shape of the object. For smooth spheres, at high Reynolds numbers, the experimental value of $C_\infty$ is around $0.47$. When particles are entrained by a turbulent flow, the same formulas remain valid, if the grain velocity $ \mathbf{v}$ is replaced by the relative velocity $\mathbf{v}-\mathbf{u}$ between the particle and the fluid velocity $\mathbf{u}$. However, the fluid flow must present negligible intrinsic turbulent fluctuations, in front of those induced by the presence of the particle.

Regarding teaching, the description of the upstream/downstream asymmetry of the flow allows students to understand that a flow at large Reynolds number $\mathcal{R}$ can either be laminar (without fluctuations) or turbulent (presenting space and time fluctuations). Conversely, at small Reynolds number $\mathcal{R}$, the flow is dominated by viscous diffusion of momentum and is therefore laminar. Moreover, turbulence leads to an increased dissipated power scaling as $\propto v^3$, to be compared to $\propto v^2$ at small Reynolds number. This constitutes a key enigma to teach: when viscosity, which is the only source of dissipation, is negligible, the generic regime is not an ideal (laminar) flow but turbulence, which leads to an enhanced dissipation.

\subsection{Molecular diffusion}
In the following, we will assume that transported particles follow the motion of the surrounding gas. Fluid and particles are then said to form an aerosol. The quantitative criteria to neglect the effect of gravity and particle inertia in front of the drag force will be derived in section~\ref{sec:inertia}. Using a continuum approach, particle velocities can be decomposed into an Eulerian velocity $\mathbf{v}$, averaged over a mesoscopic scale, and a zero-averaged random fluctuating velocity. This thermal microscopic velocity induces a diffusion of particles. If particles are initially densely concentrated near one point, they get spread out in the absence of any mean flow. Microscopically, this results from collisions with the gas molecules. In the "random walk" approximation, one can assume that the particles moves a distance $\overline l$ between collisions, separated by a time $\overline l/\overline v$, such that their direction of motion gets randomized after each collision. Although the distance $\overline l$ is actually distributed and the direction of scattering may be correlated with the direction of motion before collisions, the random walk is an effective description of diffusion at times larger than the velocity decorrelation time. After $N$ collisions, the mean distance travelled by the particle remains null but the mean squared distance is:\cite{taylor_diffusion_1922,risken}
\begin{equation}
\langle r^2\rangle=N\bar l^2=\overline l\overline v t
\end{equation}
The distance travelled typically increases as $\sqrt{t}$, which is a key characteristic of diffusion. By comparison, it increases linearly in time $t$ for a ballistic trajectory. Using the central limit theorem, the (null) average and the standard deviation at hand, the probability distribution for where a particle is after a (large) time $t$ reads:
\begin{equation}
P(r)=\sqrt{\frac{1}{2 \pi \overline l\overline vt}} \exp \left[-\frac{r^{2}}{2 t \overline l\overline v}\right]
\end{equation}

In the continuum approximation, we introduce the concentration $C$ of particles and its flux $\mathbf{j}$. The conservation equation reads:
\begin{equation}
\frac{\partial C}{\partial t}+\nabla \cdot \mathbf{j}=0
\end{equation}
Transport by the the average flow leads to a flux equal to the concentration times velocity $\mathbf{j}=C \mathbf{v}$. Consider now thermal diffusion induced by random microscopic velocity fluctuations at $\mathbf{v}=\mathbf{0}$. Would the concentration be homogeneous i.e. constant in space, the diffusive flux would vanish. At leading order, random exchanges between neighboring layers of fluid lead to a flux proportional to the gradient of concentration, oriented from high to low concentration. The phenomenological relation is known as Fick's law:
\begin{equation}
\mathbf{j}=-D \nabla C
\end{equation}
and the diffusion coefficient $D$ is assumed to be independent of $C$, which must be true in the dilute limit $C \to 0$.

Plugging this relation into the continuity equation, we get the linear diffusion equation:
\begin{equation}
\frac{\partial C}{\partial t}=D\nabla^2 C
\end{equation}
whose Green function, i.e. the exact solution for an initial condition where all the particles are concentrated at a single point reads:
\begin{equation}
C=\frac{m}{\left(4 \pi D t\right)^{3 / 2}} \exp \left[-\frac{r^{2}}{4 D t}\right]
\end{equation}
where $m$ is the total mass of particles. Identifying the microscopic model with the continuum approach, we obtain the Einstein-Smoluchowski equation:\cite{risken}
\begin{equation}
D=\frac{1}{2} \overline v \overline l.
\end{equation}

\subsection{Reynolds averaged scalar transport equation}
It has been argued in a pioneering article by Taylor~\cite{taylor_diffusion_1922} published one century ago that turbulent dispersion, induced by random turbulent fluctuations, is diffusive. Light enough particles move with the local velocity, which includes fluctuations around the average, noted $\overline v$. Their overall motion is a random walk induced by turbulent fluctuations $\mathbf{v'}$ superimposed to an averaged drift velocity $\overline{\mathbf{v}}$. This is called the Reynolds decomposition:\cite{lesieur_turbulence_2008} $\mathbf{v}=\overline{\mathbf{v}}+ {\mathbf{v'}}$. More precisely, $\overline{\mathbf{v}}$ is defined as the average of velocity over realizations of the same experiment with different turbulent fluctuations. The motion of particles is a Brownian-like motion at a scale larger than the Eulerian turbulence integral length $\mathcal{L}$, above which velocities are uncorrelated. The typical fluctuating velocity $\sigma_V$ is the root mean squared fluctuating velocity $\sigma_V^2 \equiv \overline{\mathbf{v'}^2}$.

The relevant transport equation is obtained by performing a Reynolds averaging procedure on the conservation equation governing the evolution of the concentration $C(\mathbf r, t)$. Introducing the ensemble average concentration $\overline{C}(\mathbf r, t)$, the concentration is decomposed into $C(\mathbf r, t)=\overline{C}(\mathbf r, t)+C'(\mathbf r, t)$. The evolution of $\overline{C}$ with respect to time $t$ is governed by:
\begin{equation}
\frac{\partial \overline{C}}{\partial t}+\nabla \cdot(\overline{\mathbf{v}} \overline{C})=\nabla \cdot\left(D_m \nabla \overline{C}-\overline{\mathbf{v^{\prime}} C^{\prime}}\right)
\end{equation}
where $\overline{\mathbf{v}}$ is the Reynolds averaged velocity vector, $D_m$ is the molecular diffusivity previously introduced. For the reasons invoked for molecular diffusion, $\overline{\mathbf{v^{\prime}} C^{\prime}}$ can be expressed using a gradient diffusion assumption:
\begin{equation}
\overline{\mathbf{v^{\prime}} C^{\prime}}=-D_t \nabla \overline{C}
\end{equation}
where $D_t$ is the turbulent diffusion coefficient, also called eddy diffusivity. The averaged concentration therefore obeys:
\begin{equation}
\frac{\partial \overline{C}}{\partial t}+\nabla \cdot(\overline{\mathbf{v}} \overline{C})=\nabla \cdot\left(D \nabla \overline{C}\right)
\end{equation}
where $D=D_m+D_t$ is the effective diffusion coefficient. At a time scale larger than the correlation time $\mathcal{T}$, the turbulent diffusion coefficient $D_t$ must be proportional to the product of $\mathcal{T}$ by the squared turbulent velocity $\sigma_V^2$.

For an average flow along the $x$ direction and neglecting the longitudinal diffusion, the average concentration $\overline C$ obeys a convection-diffusion equation which reduces in the steady state to:
\begin{equation}\label{eq:convection-diffusion}
\overline v \frac{\partial \overline C}{\partial x} = r^{-1} \frac{\partial}{\partial r}\left(r\;D\;\frac{\partial \overline C}{\partial r}\right)
\end{equation}
This equation has the same structure as the diffusion equation, except that time is replaced by the space coordinate $x$. In the regime described by Taylor, the turbulent diffusion coefficient $D$ does not depend on $r$, the equation admits an exact solution:
\begin{equation}
\overline C=\frac{\dot m}{\pi \sigma_R^2 \bar v}\;\exp\left(-\frac{r^2}{2\sigma_R^2}\right)
\end{equation}
where the multiplicative factor is obtained by identifying the mass flow rate across any section to the source injection rate $\dot m$. The dispersion radius $\sigma_R$ obeys the equation:
\begin{equation}
\bar v \frac{\mathrm d \sigma_R^2}{\mathrm d x}=2 D
\label{eq:radiusdiffuse}
\end{equation}
This equation can also be seen as the dispersion in time of particles injected at a point source at initial time, the Gaussian concentration profile resulting then from the central limit theorem.

Taylor's picture of turbulent diffusion is analogous to molecular diffusion, the turbulent kinetic energy playing the role of temperature. However in the turbulent regime the velocity of a fluid particle, followed along its Lagrangian path, remains correlated at short times.\cite{batchelor_diffusion_1949,batchelor_application_1950,batchelor_diffusion_1952} The velocity of a transported particle is therefore correlated as well, which leads to an anomalous diffusion at short distances. On a phenomenological ground, the simplest hypothesis would be to assume that the velocity fluctuation is a low pass filtered noise associated with a typical relaxation time ${\mathcal T}$. For a Langevin equation driven by a white noise, with a linear friction term, the correlation function decays exponentially:\cite{sawford_reynolds_1991,sawford_turbulent_2001}
\begin{equation}
\label{eq:correlation_velocity}
\overline{\mathbf{v^{\prime}}(t)\mathbf{v^{\prime}}(t+\tau)}=\sigma_V^2\exp(-\tau/{\mathcal T})
\end{equation}
The correlation time ${\mathcal T}$ is called the Lagrangian integral time scale, defined in the general case as:
\begin{equation}
{\mathcal T}=\frac{1}{\sigma_V^2}\;\int_0^\infty \overline{\mathbf{v^{\prime}}(t)\mathbf{v^{\prime}}(t+\tau)} \mathrm d\tau
\end{equation}
The dispersion of fluid particles injected at a source point at time $t=0$ is given by Taylor's theorem:\cite{taylor_diffusion_1922,sawford_turbulent_2001}
\begin{equation}
\frac{d \overline{{\mathbf r}^{2}(t)}}{d t}=2 \overline{{\mathbf r}(t) {\mathbf v}(t)}=2 \int_{0}^{t} \overline{{\mathbf v}\left(t^{\prime}\right) {\mathbf v}(t)} \mathrm d t^{\prime} .
\end{equation}
Applied to equation~(\ref{eq:correlation_velocity}) at time $t=x/\overline v$, it gives:\cite{sawford_turbulent_2001,du_estimation_1995}
\begin{equation}
\sigma_R^{2}= \frac{2}{3}\sigma_V^2 {\mathcal T}^2 \left[\exp \left(-\frac{x}{{\overline v}{\mathcal T}}\right) +\frac{x}{{\overline v}{\mathcal T}}-1\right]
\label{eq:EqDiffusionRadius}
\end{equation}
At large distance, one recovers a diffusive regime:
\begin{equation}
\sigma_R= \sigma_V \left(\frac{2 {\mathcal T} x}{3 {\overline v}}\right)^{1/2}
\label{eq:TaylorDiffusion}
\end{equation}
The concentration along the axis ($r=0$) is therefore expected to decay as $1/x$. Conversely, at short distance, the formula predicts a ballistic-like regime of the form:
\begin{equation}
\sigma_R= \frac{ \sigma_V x}{\sqrt{3} \bar v}
\label{eq:TaylorBallistic}
\end{equation}
which corresponds to a faster decay of the concentration along the axis ($r=0$) as $1/x^2$.

\subsection{Effects of inertia}
\label{sec:inertia}
Unlike passive scalars, particles are subjected to gravity and inertia. The equation of motion of a single particle reads\cite{maxey_equation_1983}
\begin{equation}
\label{eq:pdf_particule}
\frac{\mathrm d}{\mathrm d t} \mathbf r=\mathbf v, \quad \frac{\mathrm d}{\mathrm d t} \mathbf v=-\frac{1}{\tau_S}(\mathbf v-\mathbf u[\mathbf r]) + \left(1-\frac{\rho_f}{\rho_p}\right) \mathbf g
\end{equation}
where $\mathbf{u} = \overline{\mathbf{u}}+\mathbf{u'}$ is the fluid velocity and $\tau_S = \rho_p d^2/18 \eta$ the Stokes times, which is the particle response time to a change in the fluid velocity given by the Stokes force~(\ref{stokes}). In the absence of turbulence, particles fall with a velocity
\begin{equation}
\label{eq:uchute}
\mathbf v_{\rm fall}=\frac{(\rho_p - \rho_f)\mathbf g d^2}{18\eta}.
\end{equation}
If the fluid presents a constant mean velocity, particles move on the average at a velocity $\overline{\mathbf{v}} = \overline{\mathbf{u}}+\mathbf v_{\rm fall}$. Consider for simplicity that the fluid velocity correlation function decays exponentially over a time ${\mathcal T}$:
\begin{equation}
\label{eq:CorrelVelo}
\overline{\mathbf{u^{\prime}}(t)\mathbf{u^{\prime}}(t+\tau)}=\sigma_U^2 \exp(-\tau/{\mathcal T})
\end{equation}
The aerosol phase forms when turbulent velocity fluctuations are large enough to counteract particle settling,\cite{friedlander_smoke_2000} i.e. when $\sigma_U>v_{\rm fall}$.
%
%
 Performing the Reynolds decomposition $\mathbf{v}=\overline{\mathbf{v}}+ {\mathbf{v'}}$ and using the Fourier transform of the autocorrelation function, we find:
\begin{equation}
\overline{\mathbf{v^{\prime}}(t)\mathbf{v^{\prime}}(t+\tau)} = \frac{\sigma_U^2 }{1-\frac{\tau_S^2}{{\mathcal T}^2}} \left(\exp{(-\tau/\mathcal T)}-\frac{\tau_S}{{\mathcal T}}\exp{(-\tau/\tau_S)} \right)
\end{equation}
Inertia acts as a low-pass filter of the fluid velocity.\cite{bec_turbulent_2010-1} The dimensionless ratio $\mathrm{St} = \tau_S/\mathcal T$ is called the Stokes number and characterizes the relative influence of particle inertia and hydrodynamic drag. If the Stokes number $\mathrm{St}$ is much smaller than $1$, particles presents a negligible inertia and the particle velocity correlation function reduces to the fluid velocity correlation function $\overline{\mathbf{v^{\prime}}(t)\mathbf{v^{\prime}}(t+\tau)} = \overline{\mathbf{u^{\prime}}(t)\mathbf{u^{\prime}}(t+\tau)}$. Conversely, in the limit where the Stokes number $\mathrm{St}$ is much larger than $1$, the correlation function decays exponentially as $\sigma_U^2 \mathrm{St} \exp{(-\tau/\tau_S)}$.

\section{Experiments}
\label{sec:experiments}

\begin{figure}[t!]
\centering
\includegraphics[width=86mm]{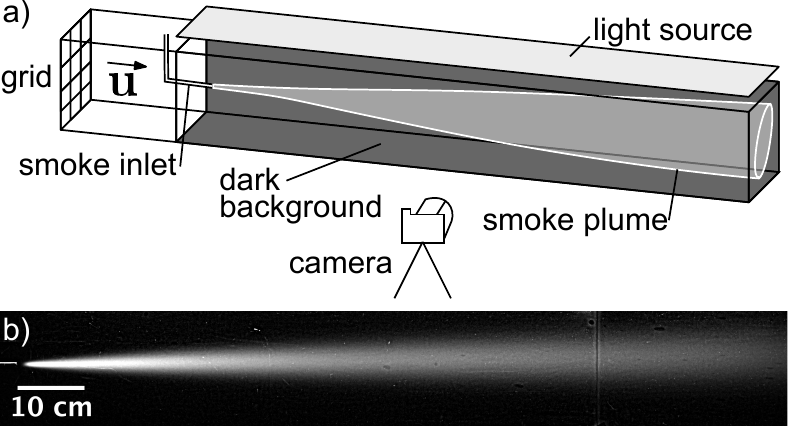}
\caption{a) Schematic of the wind tunnel. A grid generates turbulence at the entrance of the wind tunnel; it is removed to perform laminar experiments. Downstream, smoke is injected and we photograph its dispersion. b) Turbulent smoke dispersion cone for $\overline u = 15.2\;{\rm m/s}$. The injection nozzle can be seen on the left. The background intensity has been subtracted.}
\label{fig:setup_windtunnel}
\end{figure}

\subsection{Wind tunnel setup}
The dispersion of a smoke composed of micron-sized oil droplets is studied in a small wind tunnel used for experimental physics student projects, schematized in Fig.~\ref{fig:setup_windtunnel}. A turbulence generating grid ($3\times 3$ grid of $7\times 7\;{\rm cm}$ squares separated by $1\;{\rm cm}$ wide bars) is inserted immediately after the inlet contraction section. The flow Reynolds number is between $10^5$ and $10^6$. The $120 \times 23\times 23 \;{\rm cm}$ test section is illuminated from above by a LED array with a diffuser. Oil vapor is injected at a controlled rate through a $6\;{\rm mm}$ nozzle heated by a resistor at a controlled power. A smoke of droplets nucleates at few millimeters downstream. The smoke is made dilute enough so that the light intensity scattered is directly proportional to the local drop concentration. We quantitatively check the dilution by putting the light behind the tunnel rather than above. High resolution pictures with a $1$ or $2\;{\rm s}$ exposure time are taken with a Digital Single-Lens Reflex camera. In order to keep the image data linear with respect to the light intensity, the raw images files are debayered using a bilinear approximation and no further image processing is performed. A series of $10$ pictures in the same conditions is averaged to achieve a satisfying statistical convergence; the mean green channel is used to measure the light intensity. The image of the tunnel without smoke is subtracted to remove residual background light. The light intensity profile $I_0(x)$ is then calibrated using a diffusive object. It is flat along the tunnel axis, except at the start and the end of the tunnel where it falls off due to greater distance to the illumination source. The intensity ${\mathcal I}(x,y)$ of a particular pixel is an integral over the $z$ axis of the scattered intensity. The concentration profile is expected to depend on the distance to the central axis in a Gaussian way: $\sim \exp\left(-((y-y_0)^2+(z-z_0)^2)/2\sigma_R^2\right)$. Integrating it over $z$, the image intensity field ${\mathcal I}(x,y)$ remains Gaussian along $y$:
\begin{equation}
{\mathcal I}(x,y)=I_0(x) {\mathcal C}(x) \sigma_R(x) \exp{\left(-\frac{\left(y-y_0(x)\right)^2}{2\sigma_R^2(x)}\right)}
\label{eq:gauss}
\end{equation}
We therefore fit each transverse pixel line over $[y_0-6\sigma_R, y_0+6\sigma_R]$ by this corrected Gaussian profile to extract ${\mathcal C}(x)$, which is proportional to the concentration on the axis, and the radius $\sigma_R(x)$.

\begin{figure}[t!]
\centering
\includegraphics[width=86mm]{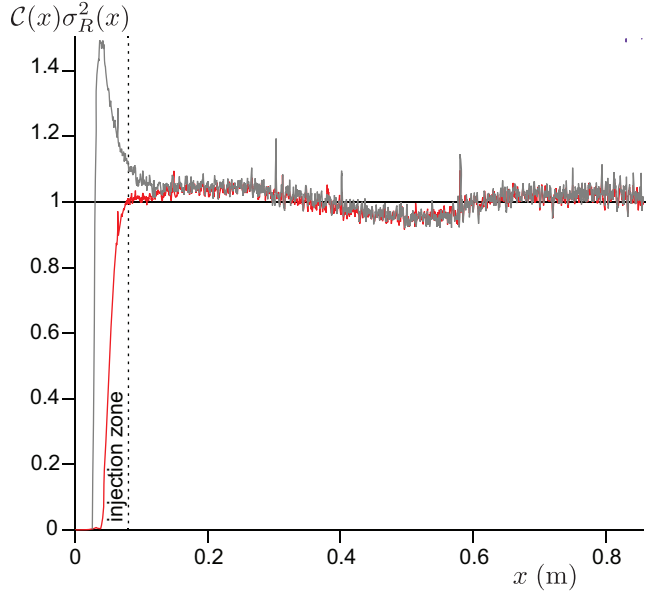}
\caption{Profiles of the rescaled smoke flow rate ${\mathcal C}(x) \sigma_R^2(x)$ as a function of distance downstream $x$. Red curve: turbulent profile for $\bar u=15\;{\rm m/s}$. Grey curve: laminar profile for $\bar u=2\;{\rm m/s}$. The two profiles collapse on the same curve except near the injection nozzle, where nucleation of oil drops from vapor takes place. The nearly constant profile shows that the droplets do not change much size during convection and dispersion. The variations, equal on both curves, are due to residual heterogeneities of light. This validates the measurement of the concentration field, within a multiplicative constant, from experimental data.}
\label{fig:ARcarre}
\end{figure}

\begin{figure}[t!]
\centering
\includegraphics[width=86mm]{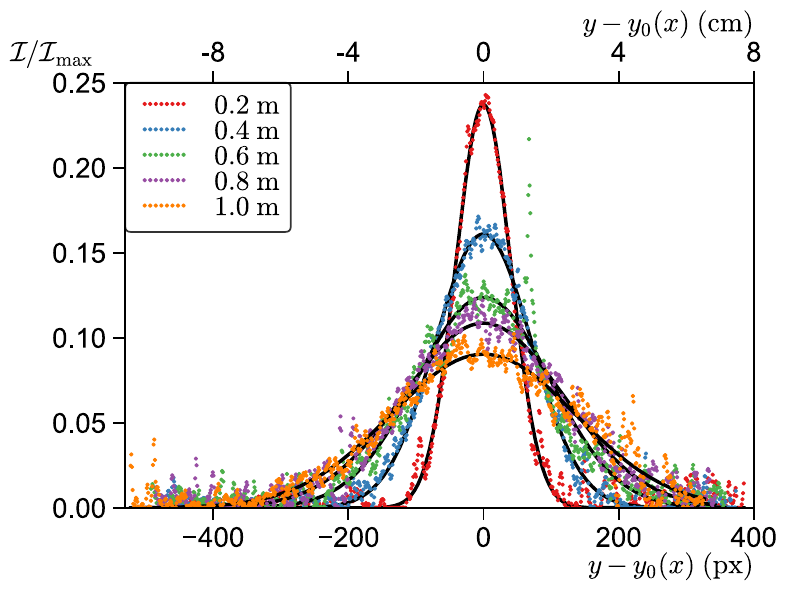}
\caption{Transverse intensity profiles ${\mathcal I}$ rescaled by the maximal image intensity at various distances from the injection nozzle. The best fit by a Gaussian is superimposed, which allows to extract the concentration ${\mathcal C}$ on the axis and the radius $\sigma_R$.}
\label{fig:gauss}
\end{figure}
\begin{figure}[t!]
\centering
\includegraphics[width=86mm]{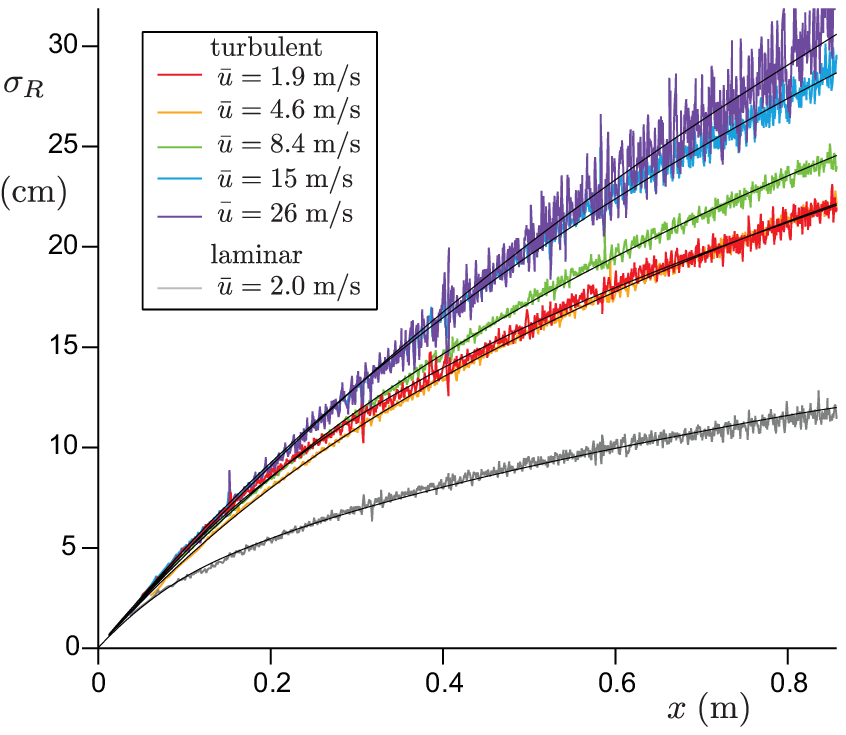}
\caption{Radius $\sigma_R(x)$ profiles as a function of distance downstream $x$. The best fit by Eq.~(\ref{eq:EqDiffusionRadius}) is superimposed, which provides measurements of the parameters $\sigma_V/\overline v$ and $\overline v \mathcal T$. The fit by a model based on a single correlation timescale is sufficient, within error bars.}
\label{fig:FigRadius}
\end{figure}

\subsection{Wind tunnel results}
Transverse intensity profiles at a regularly spaced distance from the smoke injection nozzle are shown in Fig.~\ref{fig:gauss}. As the droplets are formed by vaporizing oil, the smoke is hot and slightly buoyant so the centerline $y_0(x)$ gets slightly shifted upwards; the transverse-to-longitudinal mean particle velocity ratio is at most $10^{-2}$, meaning that the flow can be safely considered as being along the tunnel axis. The profiles are nicely fitted by a Gaussian, which leads to the measurement of the two parameters of Eq.~(\ref{eq:gauss}), the concentration on the axis, ${\mathcal C}$ and the radius $\sigma_R$. Mass conservation imposes that ${\mathcal C}(x) \sigma_R^2(x) \overline v$ must be a constant proportional to the rate of emission by the source. Fig.~\ref{fig:ARcarre} shows typical profiles ${\mathcal C}(x) \sigma_R^2(x)$ which are indeed flat outside of the injection zone. It means that the droplets, once nucleated, do not change much and keep the same light scattering properties; this validates the use of this measurement technique.

Fig.~\ref{fig:FigRadius} shows various profiles of the dispersion radius $\sigma_R(x)$. The dispersion of aerosols is significantly larger with a grid generating turbulence than without. In all cases, Eq.~(\ref{eq:EqDiffusionRadius}) provides an excellent fit to the data, within error bars. It constitutes a very striking result: a model based on a single correlation timescale is sufficient to represent adequately the data. At short distance $x<\bar v { \mathcal T}$ from the source, the dispersion takes place in a cone: $\sigma_R$ is linear in $x$. This controversial regime has been explored in a series of recent papers.\cite{villermaux_histogramme_1998,villermaux_geometry_1999,villermaux_short_2001,villermaux_mixing_2019,bec_turbulent_2010-1,bec_acceleration_2006-1,biferale_lagrangian_2008,bourgoin_role_2006,toschi_lagrangian_2009-1,bourgoin_turbulent_2015} At large distance $x>\bar v { \mathcal T}$, conversely, the constant diffusion regime predicted by Taylor is recovered.
\begin{figure}[t!]
\centering
\includegraphics[width=86mm]{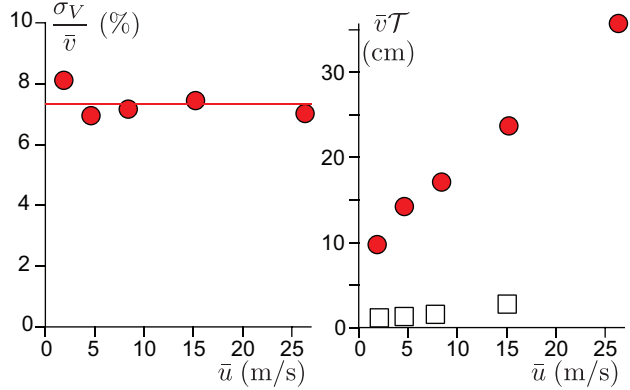}
\caption{Dependence of the fluctuation rate $\sigma_V/\overline v$ and of the cross-over length $\overline v \mathcal T$ with the mean particle velocity $\bar v$. In panel (b), the squares are measurement performed in the quasi-laminar case, in the absence of a grid at the entrance of the wind tunnel.}
\label{fig:FigLongueur}
\end{figure}

From each profile $\sigma_R(x)$, two quantities are therefore extracted: the turbulent fluctuation rate $\sigma_V/\overline v$ and the cross-over length $\overline v \mathcal T$, which is the cross-over distance between the ballistic and diffusive regimes. These quantities are plotted in Fig.~\ref{fig:FigRadius} as a function of the mean flow velocity $\overline v$. The fluctuation rate $\sigma_V/\overline v$ is independent of the wind velocity $\overline u$, as expected, and equal to $\sim 7.5\%$. The cross-over length $\overline v \mathcal T$ is very small when no grid is introduced at the entrance of the wind tunnel. It is much larger with a turbulence generating grid. It tends to a constant as $\overline u \to 0$ and increases with the fluid velocity $\overline u$. The Lagrangian correlation length $\sigma_V \mathcal T$ increases from $0.7\;{\rm cm}$ to $2.5\;{\rm cm}$ in the range of velocities explored, which is significantly smaller than the grid mesh size. The slope of the curve gives a characteristic time around $9\;{\rm ms}$, which is much larger than the Stokes time of oil droplets. To the best of our knowledge, there is no simple mechanistic explanation to the observed behaviour.

\subsection{Field experimental setup}
\begin{figure}[t!]
\centering
\includegraphics[width=86mm]{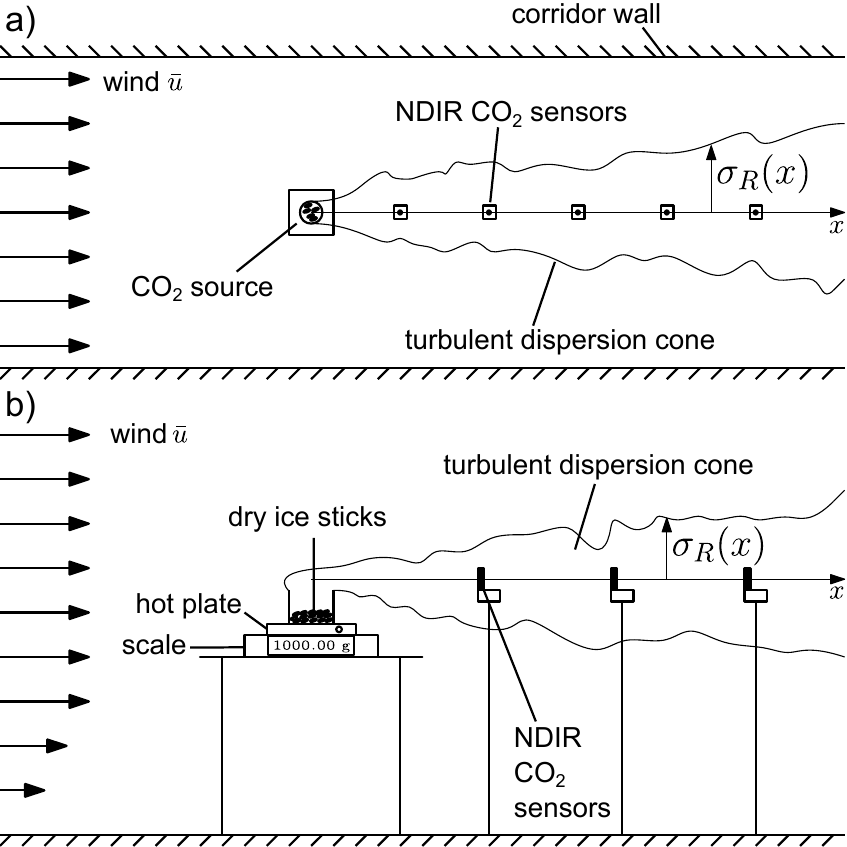}
\caption{Schematic of the experimental setup used to measure the turbulent dispersion of contaminants. Dry ice sticks stored in an open cylindrical container are sublimated using a hot plate. The gas produced is convected downstream by the small draft blowing in the corridor where the CO$_2$ concentration is recorded. The scale measures the mass injection rate $\dot m$. (a) Top view of the corridor layout. The source size ($\sim 20\;{\rm cm}$) is much smaller than the typical corridor width ($5$ to $10\;{\rm m}$). (b) Zoomed-in view of the source. The sensors are placed at $1.1 \; {\rm m}$ above ground.}
\label{fig:schema_manip}
\end{figure}
In order to test the relevance of the wind tunnel results for the transmission of SARS-CoV-2 in public spaces, we have performed "field" experiments in the large corridors of two commercial malls and of our university. The experimental set-up is schematized in Fig.~\ref{fig:schema_manip}. A controlled CO$_2$ source of constant mass rate $\dot m$ is obtained by sublimating dry ice sticks in an open cylindrical container of diameter $20 \; {\rm cm}$ heated by a power controlled hot plate. The source is positioned at a distance $1.1 \; {\rm m}$ above the ground. The imposed sublimation rate, measured using a scale, is equal $\dot{m} = 1.5\;{\rm g/s}$. This order of magnitude is chosen to measure CO$_2$ concentrations with a high relative accuracy without saturating the sensors.

Our initial motivation was to characterize the dispersion of breath in public spaces. The source used is about $150$ times the CO$_2$ exhalation rate of an adult at rest, which does not change the dispersion rate. The experiment had been designed believing that the turbulent dispersion would be statistically isotropic, as expected for an effective diffusion caused by large scale incoherent turbulent motion.\cite{cheng_modeling_2011} Preliminary observations using a source of micron sized glycerol+water droplets have shown that in most large public spaces, there are air drafts causing horizontal transport and biased dispersion. After identification of the mean air flow direction, non-dispersive infrared (NDIR) CO$_2$ sensors are placed downstream from the source. They recorded the CO$_2$ concentration $\overline C$ over the duration of each experiment, around $30$ minutes. The initial and final values of $\overline C$ are used to determine the background CO$_2$ level $C_e$. 

After an initial short transient time, a concentration field in a statistically steady state is established. CO$_2$ concentration in excess is measured in eight different locations. The draft wind velocity was measured using a hot-wire anemometer. It ranges from $0.1$ to $2\;{\rm m/s}$ depending on the location. The flow Reynolds number is between $10^5$ and $10^6$, as in the small-scale wind tunnel. Measurements are done with different ventilation flow rates and recycled air fractions; the use of fire safety ventilation, namely mechanical smoke extractors and smoke vents, has been tested when available. Measurements have been performed with entrance doors both open and closed, as open doors create large drafts in some locations. Control sensors (not shown in Fig.~\ref{fig:schema_manip}), placed immediately upstream, to the left and to the right of the source showed no concentration increase: convection dominates over turbulent diffusion.

\subsection{Field results}

\begin{figure}[t!]
\centering
\includegraphics[width=86mm]{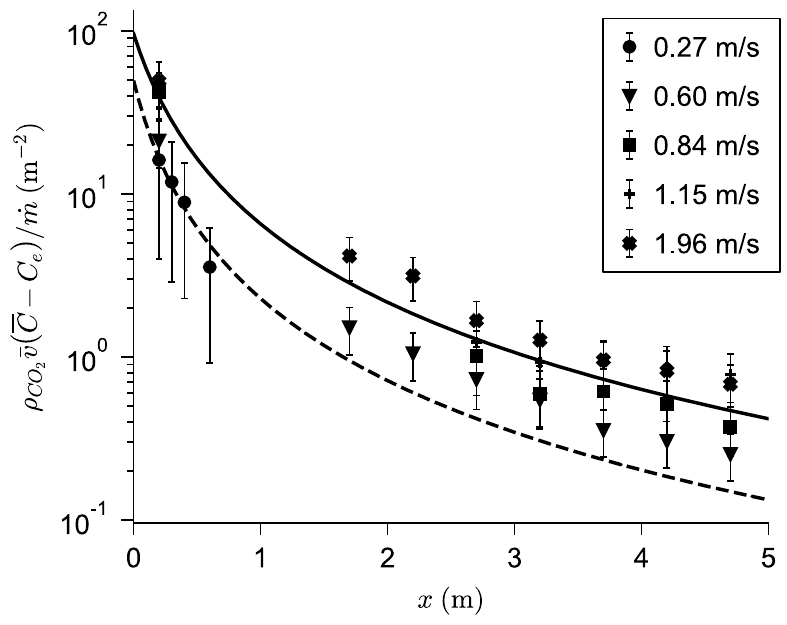}
\caption{Time-averaged concentration excess profiles as a function of distance $x$ to the source. The concentration is rescaled by $\dot{m}/ (\rho_{\mathrm{CO}_2} \overline v)$, where $\dot m$ is the mass injection rate and $\overline v$ the air velocity. The solid curve is the best fit by equation~(\ref{eq:dilution}). Disks and dashed line: human breathing measurements ($\dot m = 10\;{\rm mg/s}$) and its best fit curve.}
\label{fig:donnees_co2}
\end{figure}

Concentration profiles are averaged over the fraction of the time during which the air flow is reasonably aligned with the sensor axis. Fig.~\ref{fig:donnees_co2} shows that $\overline C$ decays spatially very fast, typically as $x^{-2}$. The mass conservation equation provides the scaling law:
\begin{equation}
\rho_{\mathrm{CO}_2}\;(\overline C-C_e)\;\overline v \sigma_R^2 \sim \dot m
\end{equation}
where $\rho_{\mathrm{CO}_2}= 1.83\;{\rm kg/m}^3$ is the density of sublimated CO$_2$. In Fig.~\ref{fig:donnees_co2}, the CO$_2$ concentration is rescaled by $\dot m/\rho_{\mathrm{CO}_2} \overline v$ and plotted as a function of space $x$. The reasonable collapse between data shows that the CO$_2$ concentration is proportional to the inverse wind velocity $\overline v$. The fraction of fresh air injected in the ventilation system had no effect on the measured concentrations: in practice, only the local airflow disperses CO$_2$. In other words, the airflow induced to renew air is negligible in front of natural air drafts.

The decay of $\overline C-C_e$ is consistent with the overall conical shape of the dispersion zone in the ballistic regime given by Eq.~(\ref{eq:TaylorBallistic}). We can parametrize the dilution by:
\begin{equation}
\frac{\rho_{\mathrm{CO}_2}\;(\overline C-C_e)\;\overline v}{\dot m} = \frac{1}{\alpha^2\;\left(a+x\right)^2}
\label{eq:dilution}
\end{equation}
where $\alpha$ is the dispersion cone slope, determined by the turbulent fluctuation rate $\sigma_V/\overline v$. We find $a = 0.35\;{\rm m}$, which is close to the actual diameter of the dry ice container and $\alpha=0.10$. We have also included in Fig.~\ref{fig:donnees_co2} the CO$_2$ concentration in the wake of a volunteer breathing through the mouth, in a fan-induced wind of velocity $\overline v=0.3\;{\rm m/s}$. We find $a \approx 0.27\;{\rm m}$ and a slightly higher fluctuation rate $\alpha=0.14$ in that case. 

\section{Relevance for the assessment of the airborne transmission risk of SARS-CoV-2}

SARS-CoV-2 is transmitted in the airborne route by respiratory aerosols:\cite{greenhalgh_ten_2021,randall_how_2021,morawska_it_2020} droplets of mucus, ranging between $200$\;{\rm nm}-$200\;{\rm \mu m}$, which can contain viral particles. These droplets are produced in the respiratory tract during normal respiratory activity by various fluid instabilities.\cite{bourouiba_fluid_2021,bourouiba_violent_2014,mittal_flow_2020} The air exhaled by an infected person contains active SARS-CoV-2 viral particles, which may start an infection when one of the virions is deposited on an ACE2 receptor of the epithelium and successfully bypasses the immune response. Infection may also start if virions deposit on the eyes. The probability of infection increases with the intake viral dose $d$, defined as the amount of virus particles inhaled by a person, cumulated over time. It therefore increases with the exposure time and with the concentration of viral particles in the air.\cite{rudnick_risk_2003,bazant_guideline_2021,poydenot_risk_2021}

The risk of viral transmission by a given infected patient is therefore determined by the concentration in viral particles in the air inhaled by another susceptible person. This concentration is given by the rate of emission of viral particles in the air exhaled by an infected patient and by the dilution of the air between exhalation and inhalation. In closed rooms, viral particles accumulate and their concentration eventually reaches a steady state between exhalation and ventilation, which is the replacement of contaminated air by fresh air. This explains the large transmission clusters in poorly ventilated indoor spaces.\cite{liu_simulation-based_2021,miller_transmission_2021} The average transmission risk, uniform in space, can be determined by assuming that viral particles are well-mixed by turbulence. In that case, dilution is due to the natural or forced air exchanges with outside air.\cite{bazant_guideline_2021,peng_exhaled_2021,rudnick_risk_2003,poydenot_risk_2021} Near an infected person however, the concentration is much higher and decays with distance, and the dilution from the exhalation concentration of viral particles to the average concentration of viral particles is controlled by the local airflow. The total transmission risk is therefore the sum of an average risk, controlled by ventilation, and a supplementary risk in the dispersion cone of exhalations. Outdoors, in the absence of homogenizing flow, this short-range risk is the only transmission risk. Respiratory activity such as unmasked coughs and sneezes can create large airflows that dominate over the preexisting flow patterns. In that case, air is exhaled in a buoyant jet that gets gradually mixed at its boundaries as it travels.\cite{bourouiba_fluid_2021,bourouiba_violent_2014}

Masks strongly alter the airflow by blocking the jet; small jets can remain along the nose and the cheeks if the mask is not properly fitted.\cite{tang_schlieren_2009,verma_visualizing_2020,deng_what_2021,viola_face_2021,feng_influence_2020} The stopping distance of a masked respiratory puff was consistently found to be\cite{deng_what_2021,verma_visualizing_2020,simha_universal_2020} around $20\;{\rm cm}$; the more filtering masks make the puff stop earlier. Beyond this very near field distance, the ambient airflow controls particle transport. The typical flow scale is therefore the size of the head, which acts as an obstacle to the flow: at a distance to the face comparable to it, the respiratory airflow has a negligible velocity, and dilution is fully controlled by the ambient airflow, which is what we have extensively developed here.

Our analysis applies to any public space where either mask wearing is mandatory and there is a horizontal air draft, or the air velocity is much larger than the exhalation velocity. We have found that in large indoor corridors, the airflow is a horizontal draft with $\overline u = 0.1-1\;{\rm m/s}$. Outdoors, the wind is also horizontal with larger velocities. Understanding the dilution in these public spaces allows us to assess their transmission risk, given biological data on pathogen contagiousness.\cite{bazant_guideline_2021,peng_exhaled_2021,rudnick_risk_2003,poydenot_risk_2021} This is needed to ground infection prevention policies on a solid physical basis, as some (such as the "6 ft rule") are not (see Refs.~\citenum{randall_how_2021},\citenum{greenhalgh_orthodoxy_2021} for a historical perspective and critical discussion). Exhaled CO$_2$ is a commonly used risk proxy in well-mixed rooms;\cite{bazant_guideline_2021,peng_exhaled_2021,rudnick_risk_2003,poydenot_risk_2021} its spatial decay allows for a simple translation of the short-range risk into an equivalent long-range risk. For the gentle draft of shopping mall corridors $\overline u\simeq 0.2\;{\rm m/s}$, the increase in risk at $6\;{\rm ft}$ corresponds to a CO$_2$ concentration excess of $50\;{\rm ppm}$.

\section{Conclusions}
\subsection{Teaching turbulence}
In this article, we have investigated experimentally and theoretically the dispersion of a passive scalar by a turbulent flow. This constitute an accessible problem to teach turbulence at an undergraduate level, combining possible experiments and theory. Using any source of smoke, a fan and a camera, the Reynolds decomposition between average and fluctuating quantities can be illustrated, qualitatively and quantitatively. The derivations, based on physical reasoning and dimensional analysis, remain linear and much simpler than attempting to solve the Navier-Stokes equation. While the Reynolds number dependence of the drag force exerted on a solid is useful to introduce dimensionless numbers, asymptotic regimes, viscosity and inertial effects, the diffusion of a passive scalar provides a way to understand the role of fluctuations to enhance turbulent mixing. The applications to the dispersion of odors and animal olfactory search,\cite{balkovsky_olfactory_2002} the dispersion of pollutants or to airborne viral transmission are immediately appealing to students.

Different concepts cannot be introduced through the dispersion of smoke, which are more difficult to understand, and in particular the concept of turbulent energy cascade. Considering a spatial scale $\ell$, the total turbulent kinetic energy can be divided into a contribution due to fluid motion at scales larger and smaller than $\ell$. This low pass and high filtering generalizes the Reynolds decomposition, introducing the scale over which the average is performed. According to the energy cascade picture, the energy is injected into large scale motion and dissipated by viscosity at small scale. In between, it must be transferred from large scales to small scales by inertial effects. The flux of kinetic energy through a particular scale $\ell$ can be defined as the transfer per unit time of kinetic energy associated to scales larger than $\ell$ into kinetic energy associated to scales smaller than $\ell$, due to velocity fluctuations of typical amplitude $\delta v(\ell)$ at scale $\ell$. Using dimensional analysis, the energy flux per unit mass across the scale $\ell$ should be proportional to $\delta v^3/\ell$. Moreover, in the range of scales for which viscous dissipation is negligible, this energy flux should be constant. One deduces the Kolmogorov scaling law:
\begin{equation}
\frac{\delta v^3}{\ell} = \frac{\sigma_V^3}{\mathcal L}
\end{equation}
where ${\mathcal L}$ is the integral scale. An analogous of the cascade idea for a passive scalar has been introduced by Richardson,\cite{richardson_atmospheric_1926,bourgoin_turbulent_2015} which would suggest an effective diffusion coefficient scaling as $\ell \delta v\sim \ell^{4/3}$ rather than proportional to $\ell$ as for the observed ballistic-like regime. The results reported here show that turbulence intermittency and coherent structures lead to a much simpler effective picture, with a single decorrelation timescale, at least for this range of Reynolds numbers.

It is important to emphasize that the equation (\ref{eq:CorrelVelo}) describes phenomenologically the velocity decorrelation in the inertial range of time-scales. Indeed, at very short time, the particle velocity can be expanded into ${\mathbf u}'(t+\tau)\simeq {\mathbf u}'(t)+{\mathbf a}(t) \tau$, where ${\mathbf a}$ is the acceleration. Therefore, at very short time-scales (say, smaller than the Kolmogorov time $(\mathcal L/\overline u) \mathcal{R}^{-1/2}$), we get:
\begin{equation}
\overline{\mathbf{u^{\prime}}(t)\mathbf{u^{\prime}}(t+\tau)}=\sigma_U^2-\frac 12 \sigma_a^2 \tau^2
\end{equation}
where $\sigma_a^2=\overline{\mathbf a^2}$. At very short time, the auto-correlation function departs quadratically from $1$ and not linearly as given by the exponential formula (\ref{eq:CorrelVelo}). This is exactly the behaviour reported both numerically and experimentally\cite{toschi_lagrangian_2009-1,biferale_lagrangian_2008} for the Lagrangian velocity structure function. It is therefore not obvious that the ballistic-like regime observed here can be interpreted as the analogous of the Batchelor regime\cite{batchelor_diffusion_1949} for the dispersion of a pair of particles.\cite{bec_turbulent_2010-1,bourgoin_role_2006,biferale_lagrangian_2008,toschi_lagrangian_2009-1,bourgoin_turbulent_2015,sawford_turbulent_2001} The Batchelor regime is indeed expected for short timescales $t<\sigma_U/\sigma_a$ and not for the inertial range of scales.

\subsection{Teaching experimental physics}
We believe that it is essential to teach experimental physics and its specific mode of reasoning. The problem of epidemic risk reduction is extremely difficult as it is at the crossroads between many different fields; it is not entirely scientific as it presents social, economic and political aspects. Yet physics must play a definite role, as it provides insights and methods to solve a few particular problems, from the biophysics of the respiratory system to infection prevention strategies.\cite{bourouiba_fluid_2021,mittal_flow_2020} Interdisciplinarity, and the contribution of the physical sciences in particular, has been shown to be essential to design an appropriate public health policy.\cite{greenhalgh_orthodoxy_2021,morawska_paradigm_2021} As such, the experiments reported in this article have been directly useful to provide rational public health policy recommendations to prevent transmission in local shopping malls on a mechanical basis. Namely: increasing turbulent dispersion or the distance between people to reduce the short-range transmission risk; increasing ventilation to reduce the long range transmission risk; performing experiments with smoke to directly test the airflows in public spaces; determining on a rational basis the limit CO$_2$ concentration compatible with a moderate airborne transmission risk. This study is based on a specific mode of reasoning, relying on experiments and measurements, rather than on theory; the mathematical tools used are simple enough to be accessible to undergraduate students.

The undergraduate students who have participated in this study have not performed (only) technical tasks: assembling NDIR CO$_2$ sensors with Arduino cards and performing measurements. They have also analyzed the problem, read parts of the literature on SARS-CoV-2, designed controlled experiments and provided practical answers. We believe that being able to read the scientific literature, to design experiments and to conduct a rigorous reasoning, is amongst the intellectual skills that the teaching of experimental physics should aim at, and not only illustrate theoretical courses with academic experiments.

\begin{acknowledgments}
Unibail-Rodamco-Westfield has funded this work under the CNRS contract 217977 and provided access to Forum des Halles and Carr\'e S\'enart as well as technical assistance. This article involves seven undergraduate students who have worked on the problem in the context of their final-year experimental physics courses ("Phy Exp") at the Universit\'e de Paris. The authors thank the "Phy Exp" team and in particular the lathe-mill operator, Wladimir Toutain, and the technician, Thibaut Fraval De Coatparquet, for their assistance. We thank Alessandra Lanotte and Luca Biferale for the fruitful discussions.

\end{acknowledgments}
\section*{conflict of interest}
This work was funded by Unibail-Rodamco-Westfield on behalf of Conseil National des Centres Commerciaux (CNCC), who asked the authors to make recommendations for a health protocol aiming to reduce and quantify the transmission risk in shopping centers. The conclusions of the present article are therefore of direct interest for the funding company. The authors declare no financial competing interest. The funding company had no such involvement in study design, in the collection, analysis, and interpretation of data, nor in the writing of the article. The authors had the full responsibility in the decision to submit it for publication.

\end{document}